\documentclass[aps,prc,twocolumn,groupedaddress]{revtex4}

\usepackage{amsmath,amssymb}
\usepackage[dvipdfm]{graphicx,color}
\usepackage{subfigure}
\usepackage{longtable}

\begin{document}


\title{Novel cluster states in $^{10}$Be}


\author{Fumiharu Kobayashi and Yoshiko Kanada-En'yo}
\affiliation{Department of Physics, Kyoto University, Kyoto 606-8502, Japan}


\date{\today}

\begin{abstract}
Cluster structures of excited $^{10}$Be states are investigated with a hybrid model of 
dineutron condensate wave functions
and $^6$He+$\alpha$ cluster wave functions. 
Two kinds of cluster states are theoretically suggested 
a few MeV above the $\alpha$+$\alpha$+$n$+$n$ threshold energy.
They have quite distinct cluster structure; 
ones have gas-like structures of $\alpha$+$\alpha$+dineutron, 
and the others have $^6$He+$\alpha$ with extremely extended an $\alpha$ cluster. 
Although these cluster states have not been confirmed experimentally yet, 
theoretically suggested properties of these states 
such as monopole transition strengths and $\alpha$-decay widths may be helpful 
for possible experimental observation.
\end{abstract}


\maketitle

\section{Introduction}
\label{Sec:introduction}

Nuclei have a variety of unique features. 
A cluster structure is one of them, 
which means that one or a few subsystems 
composed of some nucleons are formed in nuclei. 
Nowadays cluster structures have been investigated eagerly both theoretically and experimentally, 
and many efforts have been made to understand their origin and features. 
The most representative cluster is an $\alpha$ particle which is very stable 
and formed in a number of nuclear systems. 

Be isotopes are the typical systems where $\alpha$ clusters develop.
The structures of Be nuclei heavier than $^8$Be
can be described so well by a 2$\alpha$ core and 
valence neutrons around it 
\cite{seya81, oertzen96, arai96, enyo99, itagaki00, ogawa00, enyo03, ito04, ito06, suhara10}.
$^{10}$Be is the lightest even-even neutron-rich Be isotope 
and shows diverse structures composed of 2$\alpha$ plus two valence neutrons, 
such as molecular orbital structures. 
Such structures have been investigated by many theoretical methods 
such as molecular orbital models \cite{seya81, itagaki00, ito04, ito06}, 
cluster models \cite{arai96, ogawa00} 
and antisymmetrized molecular dynamics \cite{enyo99, suhara10}  and so on. 
As for its $0^+$ states, the $0^+_1$ and $0^+_2$ states have been confirmed experimentally
and their structures are well-described theoretically. 
Recently, in addition to those two $0^+$ states, 
a possibility of the existence of a new $0^+$ state was suggested experimentally
on the analogy of $^{10}$B \cite{kuchera11}. 
According to the report, the $0^+$ state should be located about $3$ MeV higher 
above the $^6$He+$\alpha$ threshold 
and might have a large $\alpha$ spectroscopic factor. 

In relation to cluster structures in nuclear physics, 
we are interested in dineutron correlation in finite nuclei.
A dineutron is composed of two neutrons coupled to spin-singlet 
which are not bound in free space. 
It is theoretically suggested that 
dineutron correlation is rather enhanced in the low-density region 
in the nuclear matter \cite{baldo90, matsuo06} 
and at the surface of finite nuclei, especially the neutron-halo 
and -skin region of neutron-rich nuclei
\cite{bertsch91, zhukov93, matsuo05, hagino05}. 
So, even if a dineutron is a fragile subsystem unlike an $\alpha$, 
it is expected that dineutron clusters are enhanced in nuclei, 
especially at the surface of neutron-rich nuclei.
It is also a challenging problem to search for developed dineutron structures 
in excited states as one suggested in $^8$He \cite{enyo07}. 
In order to investigate dineutron correlation in finite nuclei, 
we had constructed a model, 
which we name the dineutron condensate (DC) wave function, 
and applied to $^{10}$Be \cite{kobayashi11}. 
In that work, 
we suggested a new $0^+$ state composed of a 2$\alpha$ core plus a dineutron 
above the $0^+_2$ state, 
though we could not come to a conclusion about the detail of that state.
The possibility of the 2$\alpha$ core plus a dineutron structure 
has also been discussed in Ref.~\cite{itagaki08}.
Then, a question is whether the theoretically suggested 2$\alpha$ core plus a dineutron state 
in Ref.~\cite{kobayashi11} can be assigned to the experimentally suggested state 
in Ref.~\cite{kuchera11} or they are individual states above the threshold energy.

In the present work, 
in order to investigate cluster structures of $^{10}$Be in detail
from the viewpoint of 
structures of $^6$He+$\alpha$ and $\alpha$+$\alpha$+dineutron, 
we superpose two kinds of cluster wave functions, 
$^6$He+$\alpha$ cluster wave functions 
and $\alpha$+$\alpha$+$2n$ DC wave functions. 
The developed $^6$He+$\alpha$ cluster states are described in detail by
the $^6$He-$\alpha$ cluster wave functions 
and structures containing a dineutron cluster are expected to be described by the DC wave functions.
Through such a close investigation,  
we will conclude that 
the cluster states mentioned above would exist as distinct ones.
In this paper, we mainly discuss on the characteristic structures of those cluster states
which have not been established experimentally yet. 
We show that the ones have almost independent $\alpha$+$\alpha$+dineutron clusters
as suggested in our previous work, 
and that the others are the developed $^6$He+$\alpha$ cluster states, 
the $0^+$ state of which, we believe, corresponds to the state suggested 
in the experimental report of Ref.~\cite{kuchera11}.
We also discuss such properties of these states as monopole transition strengths
and $\alpha$-decay widths, 
which would be helpful information for experiments to search for these states.

The outline of this paper is as follows. 
In Sec.~\ref{Sec:framework}, we present our framework in this work briefly. 
In Sec.~\ref{Sec:result}, we show the obtained results and discuss features of the cluster states.
In Sec.~\ref{Sec:summary}, we summarize the present work.

\section{Framework}
\label{Sec:framework}

In the present work, 
we superpose two kinds of cluster wave functions in order to describe $^{10}$Be states, 
$^6$He+$\alpha$ cluster wave functions used in Ref.~\cite{enyo12} 
and $^{10}$Be DC wave functions used in Ref.~\cite{kobayashi11}. 
The details of the formalism for each model are explained in these references 
and we do not enter the details here.

\subsection{The $^6$He+$\alpha$ cluster wave function}
\label{Sec:cluster_wf}
In $^6$He+$\alpha$ cluster wave functions,  
$|\Phi_{{\rm c}}(d_c) \rangle$, 
the $^6$He and $\alpha$ clusters are separated by a distance $d_c$. 
The spatial parts of the single-particle wave functions in $\alpha$ and $^6$He are described 
by the harmonic oscillator (H.O.) shell model wave functions whose centers of mass are 
located at $(0,0,3d_c/5)$ and $(0,0,-2d_c/5)$ respectively
and widths are the same value of $\nu=1/(2b^2)=0.235$ fm$^{-2}$. 
In an $\alpha$ cluster, 
four nucleons are in the $s$-shell closed $\pi[(0s)^2]\nu[(0s)^2]$.  
The $^6$He cluster is composed of an $\alpha$ core 
and two valence neutrons in $0p$-orbits around the $\alpha$ core. 
We consider all the $0p$-shell configurations for two valence neutrons
and mix those configurations to describe the $^6$He cluster. 
We describe the $0p$-configurations with the shifted Gaussians from the $\alpha$ core 
as in Ref.~\cite{enyo12}. 
An isolate $^6$He described in such the simple configurations of the H.O. shell model 
does not necessarily agree with the realistic one qualitatively. 
For example, its binding energy is higher than 
the threshold energy to $\alpha$+$n$+$n$
and it has no neutron-halo structure. 
In spite of such a simplified $^6$He,
the structure of the well-known states of $^{10}$Be can be described well 
with the $^6$He+$\alpha$ cluster wave functions
as discussed in Ref.~\cite{enyo12}.

\subsection{The $^{10}$Be DC wave function}
\label{Sec:DC_wf}
In addition to the $^6$He+$\alpha$ cluster wave functions, 
we superpose the DC wave functions containing dineutron correlation explicitly. 
The $^{10}$Be DC wave function is composed of 
the determinant of the eight single-particle wave functions 
in the 2$\alpha$ core and the two ones in a dineutron
which contains two neutrons coupled to spin-singlet. 
Its spatial form is
\begin{align}
\Phi_{\rm DC} = &\ \frac{1}{\sqrt{10!}} \int d^3 \boldsymbol{Y}_n
\exp \left[ - \left( \frac{\boldsymbol{Y}_n}{B_n} \right)^2 \right] 
\nonumber \\
&\ \times \det \left[  \psi_{\alpha 1}(\boldsymbol{r}_1) \cdots \psi_{\alpha 1}(\boldsymbol{r}_4) \right. \nonumber \\
&\ \left. \hspace{2em}
\times \psi_{\alpha 2}(\boldsymbol{r}_5) \cdots \psi_{\alpha 2}(\boldsymbol{r}_8) \ \psi_n(\boldsymbol{r}_{n1}) \psi_n(\boldsymbol{r}_{n2}) \right], \label{eq:DCwf} \\
\psi_n(\boldsymbol{r}) &\ \equiv \left( \frac{1}{\pi b_n^2} \right) ^{3/4} 
\exp \left[ -\frac{1}{2b_n^2} \left( \boldsymbol{r} - \boldsymbol{Y}_n \right)^2 \right]. \label{eq:DCwf_n}
\end{align}
In Eq.~(\ref{eq:DCwf}), the single-particle wave functions in an $\alpha$ cluster, 
$\psi_{\alpha i} \ (i = 1, 2)$,  
are the same as used in $^6$He+$\alpha$ cluster wave functions, 
whose centers are at $(0,0,\pm d_n/2)$
and widths are $0.235$ fm$^{-2}$. 
The wave functions of two neutrons in the dineutron have 
the common Gaussian width, $b_n$, and center, $\boldsymbol{Y}_n$. 
The center $\boldsymbol{Y}_n$ is integrated out multiplied by the Gaussian weight.  
The width of the Gaussian weight is $B_n$
and the dineutron is distributed around the 2$\alpha$ core 
in the spherical $S$-orbit on the whole space. 
In order to make clear the relative motion between two neutrons 
and between the dineutron and the core, 
we perform the $\boldsymbol{Y}_n$-integral and rewrite two neutron wave functions 
to the relative and center of mass wave functions of two neutrons as follows. 
\begin{align}
& \Phi_{\rm DC} = \frac{1}{\sqrt{10!}} 
\det \left[ \psi_{\alpha 1}(\boldsymbol{r}_1) \cdots 
\psi_{\alpha 1}(\boldsymbol{r}_4) \right. \nonumber \\ 
& \left. \hspace{6.5em} \times \psi_{\alpha 2}(\boldsymbol{r}_5) \cdots 
\psi_{\alpha 2}(\boldsymbol{r}_8) 
\ \psi_r(\boldsymbol{r}) \psi_G(\boldsymbol{r}_{G}) \right], \label{eq:DCwf2} \\
& \psi_r (\boldsymbol{r}) \equiv \left( \frac{B_n}{b_n} \right)^3 
\exp \left[ - \frac{r^2}{4b_n^2} \right],
\hspace{1em}
\boldsymbol{r} = \boldsymbol{r}_{n1} - \boldsymbol{r}_{n2}, \label{eq:DCwf_rel} \\
& \psi_G (\boldsymbol{r}_G)  \equiv \exp \left[ - \frac{r_G^2}{\beta^2} \right],
\hspace{4em}
\boldsymbol{r}_G = \frac{\boldsymbol{r}_{n1} + \boldsymbol{r}_{n2}}{2}. 
\label{eq:DCwf_com}
\end{align}
Here we define a new parameter $\beta^2 \equiv B_n^2 + b_n^2$. 
The parameter $b_n$ is the width of the relative wave function 
(Eq.~(\ref{eq:DCwf_rel})), 
and $\beta$ stands for the one of the dineutron center of mass wave function (Eq.~(\ref{eq:DCwf_com})) 
so that we call the parameters $b_n$ and $\beta$ 
as the dineutron size 
and the dineutron extension from the core respectively. 
In DC wave functions, we take various values for the parameters $b_n$ and $\beta$ 
and superpose the DC wave functions
to describe the changing effect of the dineutron structure 
as well as the one of the dineutron size. 
It should be noted that the Gaussian width for two neutrons in a dineutron, $b_n$, 
is generally different from the one of the single-particle wave functions 
in the core, $b$, and as a consequence, 
it is difficult to separate the center of mass motion exactly. 
So, in calculating energy expectation values, 
we approximately treat the center of mass motion effect 
by extracting the expectation value of center of mass kinetic energy from the total energy. 
On the other hand, in calculating root mean square radii and so on, 
we neglect the effects of the center of mass motion for simplicity.

\subsection{The $^{10}$Be wave function in the present work}
\label{Sec:c+DC_wf}
By superposing the $^6$He+$\alpha$ cluster wave functions 
and $^{10}$Be DC wave functions 
and projecting each wave function to the parity and angular momentum eigenstates, 
we describe $^{10}$Be$(J^{\pi})$ states, $|\Psi_M^{J \pi} \rangle$.
\begin{equation}
|\Psi_M^{J \pi} \rangle = \sum_K \left( \sum_i c_{Ki} 
\mathcal{P}^{J \pi}_{MK} |\Phi_{{\rm c},i} \rangle 
+ \sum_j c_{Kj} \mathcal{P}^{J \pi}_{MK} |\Phi_{{\rm DC},j} \rangle \right). 
\label{eq:Be10wf}
\end{equation}
Here, the indices $i,j$ are the abbreviations of a set of the parameters,
that is, $i = \{ d_c \}$ and $j = \{ d_n, b_n, \beta \}$, 
and the coefficients $c_{Ki,Kj}$ are determined by diagonalizing the Hamiltonian.
In the present work, 
we superpose the $^6$He+$\alpha$ cluster wave functions of $d_c=1, 2, \cdots 8$ fm
to describe the structure of $^6$He plus a rather developed $\alpha$ cluster. 
We also superpose the DC wave functions of five $b_n$ values, 
$\beta=2, 3, \cdots 9$ fm and $d_n=1, 2, \cdots 6$ fm. 
The set of $b_n$ values are chosen for each $\beta$
in the same way as in Ref.~\cite{kobayashi11}. 
We superpose DC wave functions with rather large $\beta$ (up to $9$ fm) 
in order to describe the state of interest which contains a weakly interacting dineutron
with respect to two $\alpha$s. 
Such constraints correspond to a bound-state approximation 
with respect to $\alpha$+$\alpha$+$n$+$n$ and $^6$He+$\alpha$.

In the present work, we investigate only the even-parity states 
because the DC wave functions of 2$\alpha$ plus a dineutron contain only even-parity components.

\section{Results}
\label{Sec:result}

\subsection{The effective Hamiltonian}
\label{Sec:hamiltonian}
In this work, we use the effective Hamiltonian as
\begin{equation}
H = T - T_G + V_{\rm cent} + V_{\rm LS} + V_{\rm Coul}, 
\label{eq:hamiltonian}
\end{equation}
where $T$ and $T_G$ are the total and center of mass kinetic energy. 
We used the Volkov No.2 force \cite{volkov65} as the central force, $V_{\rm cent}$, 
and the spin-orbit part in the G3RS force \cite{tamagaki68} 
as the spin-orbit force, $V_{\rm LS}$. 
The Coulomb force, $V_{\rm Coul}$, is approximated by seven-range Gaussians.
The adopted parameters are the same as those used
in Refs.~\cite{suhara10, kobayashi11, enyo12}, 
that is, in the central force $m=0.60$, $b=h=0.125$ 
and in the spin-orbit force the strength $v_1 = -v_2 = 1600$ MeV. 
We also use modified parameters below
to see the dependence of the level structure of the states of interest 
on the parameter choice.

\subsection{Energy spectra} 
\label{Sec:spectrum}
\begin{figure}[b]
\includegraphics[scale=0.65]{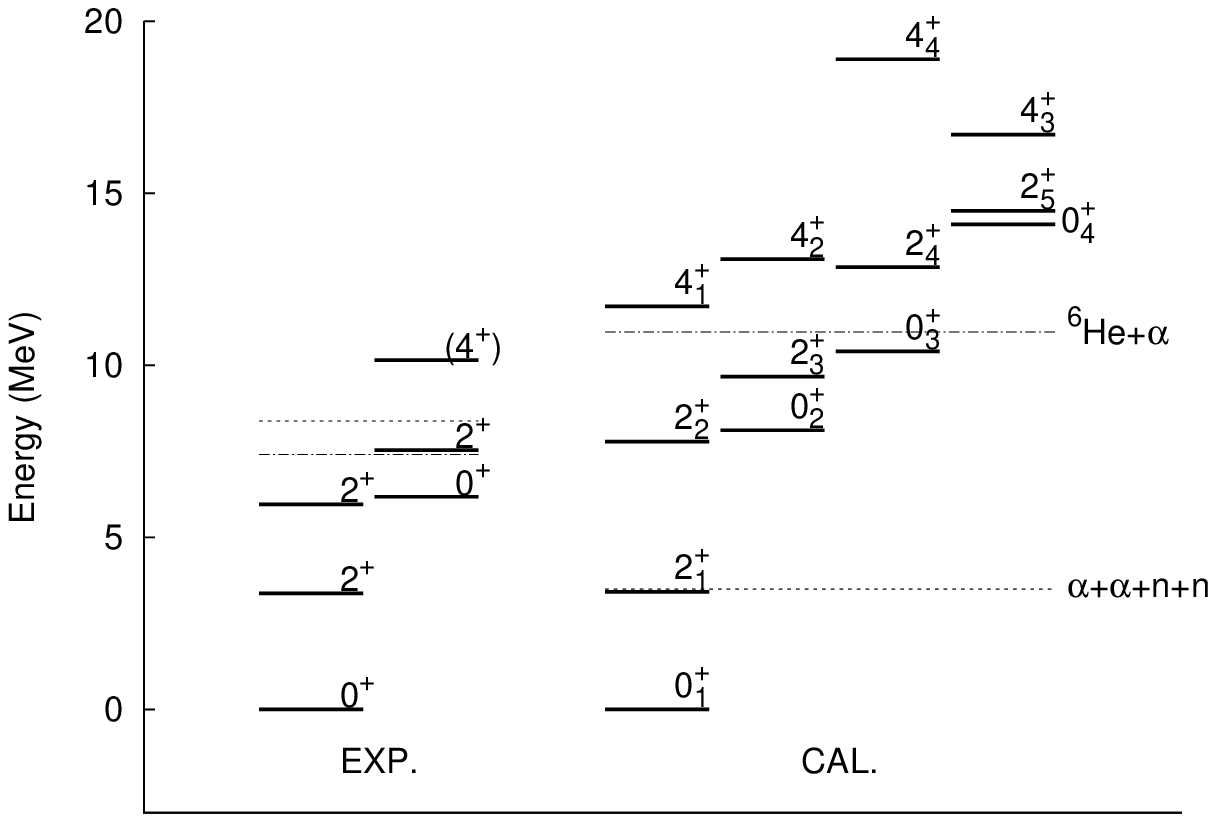}
\caption{Energy spectra of experiments and the present calculation 
on the basis of the ground state energy. 
The experimental and calculated threshold energy to $\alpha$+$\alpha$+$n$+$n$ 
and to $^6$He+$\alpha$ 
are shown with dotted and dashed-dotted lines respectively.
The assigned spin-parity, $J^{\pi}$, are labeled to each state 
and the orders only in our result. }
\label{fig:ene_spect}
\end{figure}

The energy spectra of the even-parity states obtained by the present calculation 
compared with experimental one are shown in FIG.~\ref{fig:ene_spect}. 
Experimentally established $0^+$ states are the ground $0^+$ state 
and the first excited $0^+$ state at $E_x = 6.2$ MeV, 
which exist  $8.4$ MeV and $2.2$ MeV 
below the $\alpha$+$\alpha$+$n$+$n$ threshold, respectively. 
The $0^+_{1,2}$ states obtained in the present results correspond to these two $0^+$ states.
In the calculation, 
the energy positions of the $0^+_1$ and $0^+_2$ states 
relative to the $\alpha$+$\alpha$+$n$+$n$ threshold energy 
are overestimated compared to the experimental values. 
This discrepancy mainly originates in the fact that in the present calculation
the energy of $^6$He is overestimated
and the $^6$He+$\alpha$ threshold is much higher than 
the $\alpha$+$\alpha$+$n$+$n$ threshold in contradiction to experimental data.
Such a disagreement is also seen in other calculations 
with similar effective interactions 
\cite{itagaki00, enyo99, suhara10}.  
In spite of the failure in the quantitative reproduction of the energy position, 
the energy levels and structure properties of the states in $^{10}$Be 
have been reasonably reproduced by the pioneering works. 
As shown later, the disagreement can be improved by slight modification of interaction parameters, 
which gives qualitatively similar results in the present calculation.

In addition to the $0^+_{1,2}$ states, 
two $0^+$ states above the $0^+_2$ state are obtained in the present calculation, 
which we tentatively label $0^+_{3,4}$ states here. 
Each of them forms the rotational band with the members of $0^+, 2^+$ and $4^+$ states. 
In the $0^+_3$ band, 
the angular momentum $J$ is produced by the rotation of 2$\alpha$ relative motion, 
while in the $0^+_4$ one, 
it is mainly contributed by the angular momentum of $^6$He and $\alpha$ relative motion.
The members have the same structure with those of their band head states, 
so we focus on the $0^+_{3,4}$ states in this paper.  
As shown closely in the next subsection, 
the $0^+_3$ state is mainly composed of developed two $\alpha$s 
and one dineutron cluster 
and the $0^+_4$ state has mainly $^6$He($0^+$) with an extremely developed $\alpha$. 
Since the $0^+_3$ and $0^+_4$ states are expected to be quasi-bound states, 
they should be coupled to continuum states of 
$\alpha$+$\alpha$+$n$+$n$ and $^6$He+$\alpha$ respectively. 
In fact, although these states are calculated within a bound-state approximation 
as mentioned in the previous section, 
they are still coupled to the neighboring continuum states to some extent
and the continuum states have some fragments of the $0^+_{3,4}$ states. 
In the present paper, we consider the states having the most remarkable cluster structures 
which we show in the next subsection
as each represent of $0^+_{3,4}$ without a width for simplicity. 
In addition to considering the significant cluster components in each state, 
we also check the identification of those quasi-bound state 
by using the pseudo potential and separating from continuum states as explained in Appendix. 

\begin{figure}[t]
\includegraphics[scale=0.65]{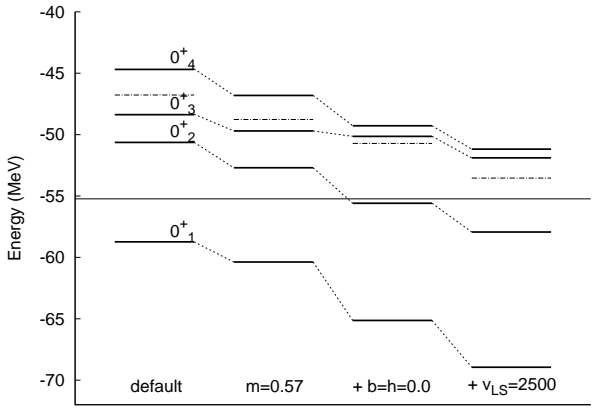}
\caption{Energy spectra calculated with different parameter sets. 
Each set is explained in the text.
The correspondent states are linked with dotted lines.
And the threshold energy to $\alpha$+$\alpha$+$n$+$n$, a constant of $-55.23$ MeV, 
and the ones to $^6$He+$\alpha$, which depends on the parameters, 
are shown with a thin solid and dash-dotted lines respectively.}
\label{fig:ene_spect_para}
\end{figure}
Here it should be noticed that the $^6$He+$\alpha$ threshold is much higher 
than the $\alpha$+$\alpha$+$n$+$n$ threshold in the present calculation, 
as is different from the experimental fact that they almost degenerate. 
The failure originates in that 
it is generally difficult to reproduce systematically the binding energies of subsystems 
by effective two-body nuclear interactions. 
In this work, we focus on the $^6$He+$\alpha$ cluster states 
and the 2$\alpha$ and a dineutron cluster states, 
which might be sensitive to the corresponding threshold energies, 
and therefore, the relative position of these threshold energies can be important 
to conclude the level structure of the $0^+_{3,4}$ states. 
To ensure the interaction dependence of the results, 
we calculate the $0^+$ energies 
by using four kinds of different sets of parameters. 
The energy spectra with the different parameter sets are plotted
in FIG.~\ref{fig:ene_spect_para}. 
In the first column which is labeled as ``default'', we plot the spectrum calculated 
with the original parameter set of $m = 0.60$ and $b = h = 0.125$ in the Volkov No.2 force 
and $v_1 = - v_2 = 1600$ MeV as the strength in the spin-orbit force, 
as mentioned in Sec.~\ref{Sec:hamiltonian}. 
In the second labeled ``m=0.57'', 
we calculate by changing only the strength of Majorana exchange term to $m = 0.57$. 
The spectrum with $b = h = 0.0$ for the strengths of Bartlett and Heisenberg terms
and $m = 0.57$ is shown in the third column (+ b=h=0.0). 
Besides the set ($b=h=0$, $m=0.57$) in the third, 
we modify the strength of the spin-orbit force to $2500$ MeV 
and show the calculated spectrum in the last column (+v$_{\rm LS}$=2500). 
By such a series of the change in the parameters, 
the binding energy of $^6$He decreases but the $\alpha$ energy is unchanged, 
so that the $^6$He+$\alpha$ threshold position relative to the $\alpha$+$\alpha$+$n$+$n$ 
threshold is improved substantially. 
It is found that the level ordering is qualitatively similar for these four parameter choices. 
In particular, the energy positions of the $0^+_2$ and $0^+_4$ states 
relative to the $^6$He+$\alpha$ energy is not sensitive to the parameter choice. 
On the other hand, the $0^+_3$ state, which have a quite distinct component 
from those of the $0^+_{2,4}$ states, shows a little different dependence. 
So it would be possible that, as shown in the fourth column,  
the $0^+_{3,4}$ states are located at closer energy positions, 
that is, a few MeV above the threshold to $\alpha$+$\alpha$+$n$+$n$ or to $^6$He+$\alpha$. 

In spite of the interaction dependence of the energies, 
the qualitative features of the structure of those states are almost unchanged.
Below, we discuss the results obtained by the original interaction set.

\subsection{Structures of the $0^+_{3,4}$ states}
\label{Sec:structure}
We investigate the detailed structures of the $0^+_{3,4}$ states, 
which we suggest are characterized by developed two $\alpha$s and one dineutron clusters, 
and extremely separate $^6$He and $\alpha$ clusters, respectively.

\subsubsection{The $0^+_3$ state of two $\alpha$s plus one dineutron}
\label{Sec:0+_3}
\begin{figure*}[t]
\begin{tabular}{cc}
\vspace{-2em}
{\includegraphics[scale=0.65]{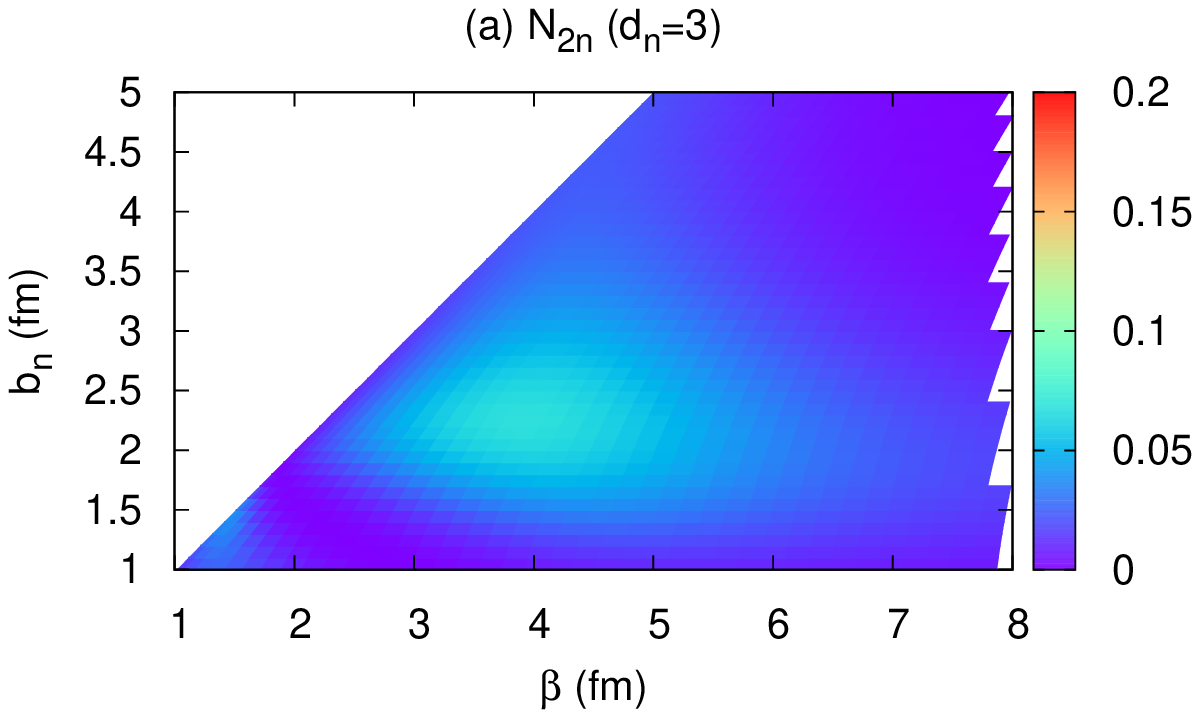}} &
{\includegraphics[scale=0.65]{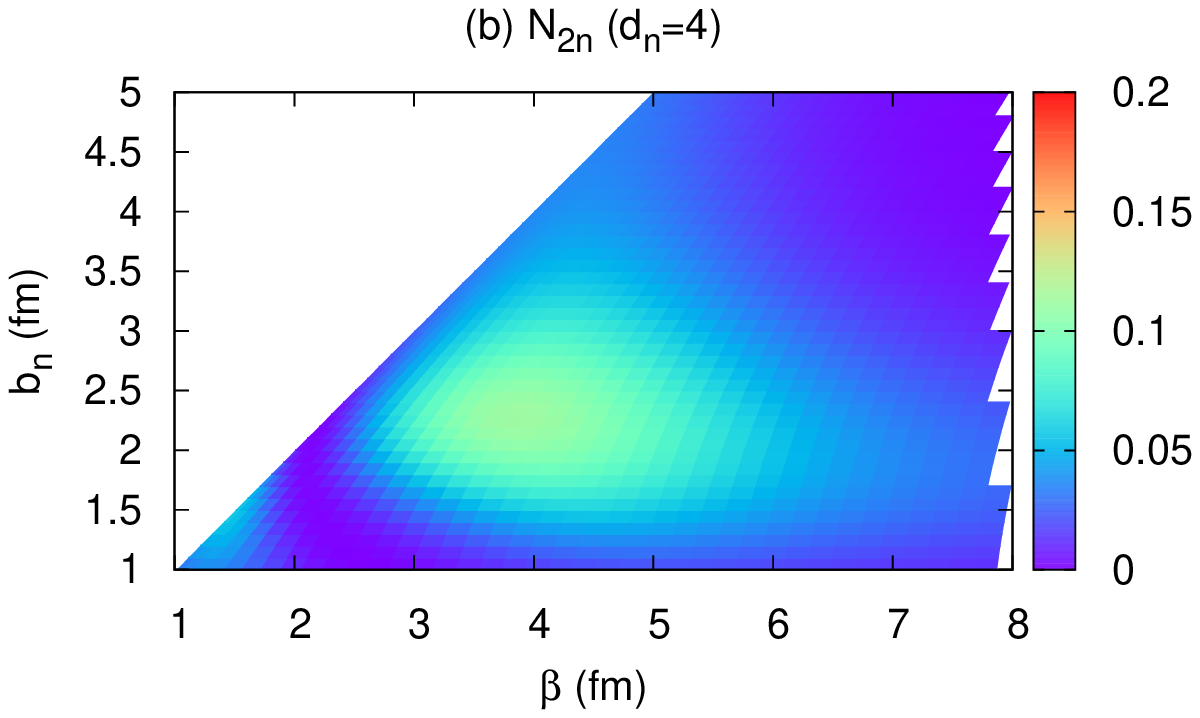}} \\
\vspace{-2em}
{\includegraphics[scale=0.65]{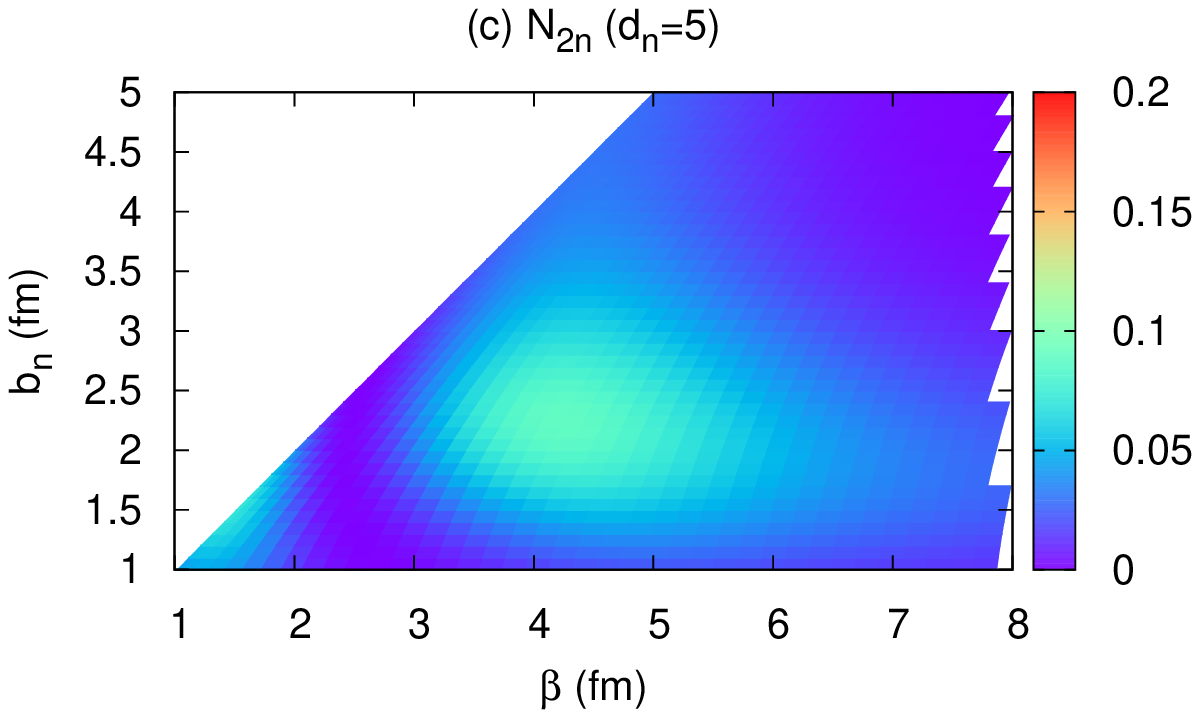}} &
{\includegraphics[scale=0.65]{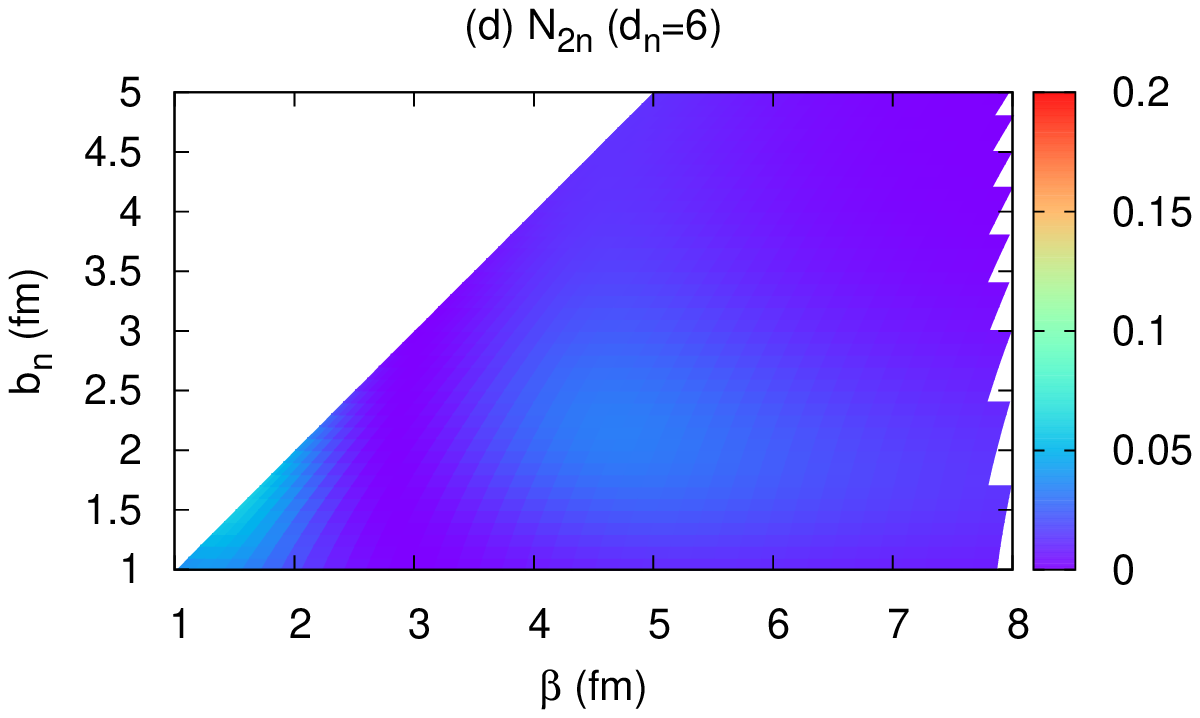}} \\
\end{tabular}
\caption{The overlaps of the $0^+_3$ state with DC wave functions, 
$\mathcal{N}_{2n}(b_n,\beta,d_n)$,
whose $d_n$ is fixed to $3 - 6$ fm ((a) $-$ (d)). 
The horizontal axis is the spread of the dineutron from the core, $\beta$, 
and the vertical axis is the size of the dineutron, $b_n$.}
\label{fig:overlap_B_b}
\end{figure*}

At first, we discuss the dineutron component in the $0^+_3$ state. 
We show the overlaps with DC wave functions, $\mathcal{N}_{2n}$, 
as the components of a dineutron cluster, 
\begin{equation}
\mathcal{N}_{2n}(b_n,\beta,d_n) 
\equiv |\langle \Phi_{\rm DC}^{0^+}(b_n,\beta,d_n)
|\Psi^{0^+} \rangle|^2, 
\end{equation}
plotted on the $\beta$-$b_n$ plane for fixed values of $d_n$ in FIG.~\ref{fig:overlap_B_b}.
A component with a small $b_n$ corresponds to the one of a compact dineutron, 
and a component with large $\beta$ 
to the one of two neutrons distributed far from the 2$\alpha$ core, 
and {\it vice versa}. 
The significant broad peak is seen at the region of 
$3 \lesssim \beta \lesssim 6$ and $1.5 \lesssim b_n \lesssim 3$.
These major peaks correspond to the component of a rather compact dineutron 
far from the core. 
The significant amplitudes of the peaks existing in the $d_n = 3,4,5$ fm cases indicate that
the $\alpha$-$\alpha$ distance fluctuates in the $0^+_3$ state. 
The minor peak around $(\beta, b_n) \sim (1.5, 1.5)$ corresponds to 
the one of two neutrons distributed in the middle of two distant $\alpha$s. 
Since such the $(\beta,b_n)  \sim (1.5, 1.5)$ component is the major component in the $0^+_2$ state 
as shown in Ref.~\cite{kobayashi11}, 
the amplitudes around the minor peak in the $0^+_3$ state are thought to come from 
the orthogonality to the $0^+_2$. 
These behaviors suggest that two $\alpha$ and one dineutron clusters 
are weakly interacting to each other and spread over a broad space, 
that is, behave as a gas-like state consisting of two $\alpha$ and one dineutron clusters. 

We should offer a comment on a difference between 
the behavior of a dineutron in the $0^+_1$ and $0^+_3$ states. 
As discussed in Ref.~\cite{kobayashi11}, 
the $0^+_1$ state has components of a dineutron 
whose peak is at $(\beta, b_n) \sim (2.0, 1.5)$, 
which corresponds to a very compact dineutron just at the nuclear surface. 
On the other hand, 
the $0^+_3$ state has a main peak at $(\beta,b_n) \sim (4.0, 2.3)$ in FIG.~\ref{fig:overlap_B_b}(b),  
where a dineutron has a larger size and a broader expansion from the 2$\alpha$ core 
than those in the $0^+_1$ state. 
We suppose it can be said that 
a dineutron is so fragile that its size is very changeable 
according to the very nuclear structure. 

As mentioned above, the $^{10}$Be($0^+_3$) shows 
the behaviors of weakly interacting two $\alpha$ and one dineutron clusters, 
which can be associated with the $3\alpha$-cluster gas state suggested 
in the Hoyle state, $^{12}$C($0^+_2$), 
by replacing one of three $\alpha$s with a dineutron cluster. 
However, it is expected that the dineutron cluster is not as rigid as an $\alpha$ 
because, as well known, two neutrons are not bound in free space 
and even in nuclear media a dineutron is naturally expected to be 
a rather weakly (semi)bound subsystem. 
In particular, the dineutron size may enlarge as the dineutron goes away from the core.
Indeed, the representative $b_n$ value at the peaks, $\sim 2.3$ fm, is 
larger than the size of an $\alpha$, $\sim 1.5$ fm. 
Furthermore, it is seen in FIG.~\ref{fig:overlap_B_b} 
that the tail part of the main peak spreads to large $\beta$ region. 
It suggests that two neutrons would expand broader and 
the $^{10}$Be($0^+_3$) may have a broad width inevitably.

\subsubsection{The $0^+_4$ state of $^6$He plus $\alpha$}
\label{Sec:0+_4}
\begin{figure}[b]
\includegraphics[scale=0.65]{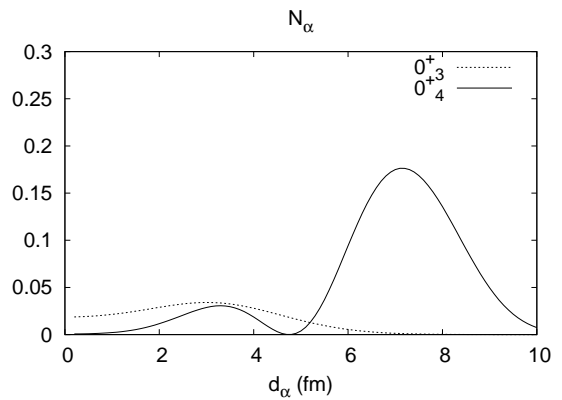}
\caption{The overlaps of the $0^+_{3,4}$ states 
with $^6$He+$\alpha$ cluster wave functions, $\mathcal{N}_{\alpha}$,
which are plotted with a dotted and solid line respectively.
The horizontal axis is the distance between $^6$He and $\alpha$.}
\label{fig:overlap_cluster}
\end{figure}

Next, we discuss $^6$He+$\alpha$ cluster structure of the $0^+_{4}$ state. 
To see how far an $\alpha$ cluster is separated spatially  from $^6$He, 
we analyze the overlap with $^6$He+$\alpha$ cluster wave functions, 
\begin{equation}
\mathcal{N}_{\alpha} (d_{\alpha}) \equiv |\langle \mathcal{P}^{0+} 
\Phi_{^6{\rm He}(0^+_1)} \otimes \Phi_{\alpha} (d_{\alpha})|\Psi^{0^+} \rangle|^2.
\end{equation}
Here the wave function $\Phi_{^6{\rm He}, \alpha}$ is fixed 
at the cluster distance $d_{\alpha}$
and the total wave function is normalized and projected to the $0^+$ state. 
The $^6$He ground state wave function $\Phi_{^6{\rm He}(0^+_1)}$ is approximated 
to be the lowest state calculated in the H.O. $p$-shell configurations 
by using the present effective interactions. 
When $d_{\alpha}$ is large, 
it corresponds to the components of a developed $\alpha$ cluster 
plus the ground state of $^6$He($0^+$).

The overlaps of the $0^+_{3,4}$ states are shown in FIG.~\ref{fig:overlap_cluster}. 
It is found that the $0^+_4$ state has striking amplitude 
in the large $d_{\alpha}$ region,
whose peak is at about $7$ fm. 
It indicates an extremely developed $\alpha$ cluster in the $0^+_4$ state, 
and that is consistent with the experimental indication in Ref.~\cite{kuchera11}. 
Such a large distance as $7$ fm is almost out of the inter-nucleus potential range.
In back of the mechanism of the formation of the $0^+_4$ state, 
a rather low Coulomb barrier exists around $8$ fm from $^6$He. 
The $\alpha$ cluster is confined in such a wide region
and constitutes the so-called ``$\alpha$-halo'' referred in Ref.~\cite{funaki05}. 
This state should be measured by $\alpha$-decay 
due to such a developed $\alpha$ cluster. 

In contrast to the case of the $0^+_4$ state, the $0^+_3$ state shows not significant
but only tiny amplitude in the $d_{\alpha} \lesssim 6$ fm region.
This state has developed two $\alpha$s and one dineutron. 
In such the state with the weak correlation between $\alpha$s and a dineutron, 
a $^6$He cluster cannot be formed. 
It can be a reason why the component of a single separated $\alpha$ from the $^6$He cluster is 
suppressed in the $0^+_3$ state. 

More detailed analysis of inter-cluster motion between $^6$He and $\alpha$ clusters 
will be given in the next subsection.

\subsection{Discussion on experimental observables}
\label{Sec:observation}
As mentioned above, these $0^+_{3,4}$ states have distinct structures from each other, 
and it is expected that 
they can be distinguished in experiments by measuring characteristic quantities.
Here we discuss the $\alpha$-decay widths and the monopole transition strengths
for the $0^+_{3,4}$ states, 
and show that the $\alpha$-decay would be useful to confirm the $0^+_4$ state 
and that the monopole transition is promising to observe the $0^+_3$ state 
in inelastic scattering experiments. 

\subsubsection{The $\alpha$-decay width}
\label{Sec:alpha_decay}
\begin{figure}[b]
{\includegraphics[scale=0.65]{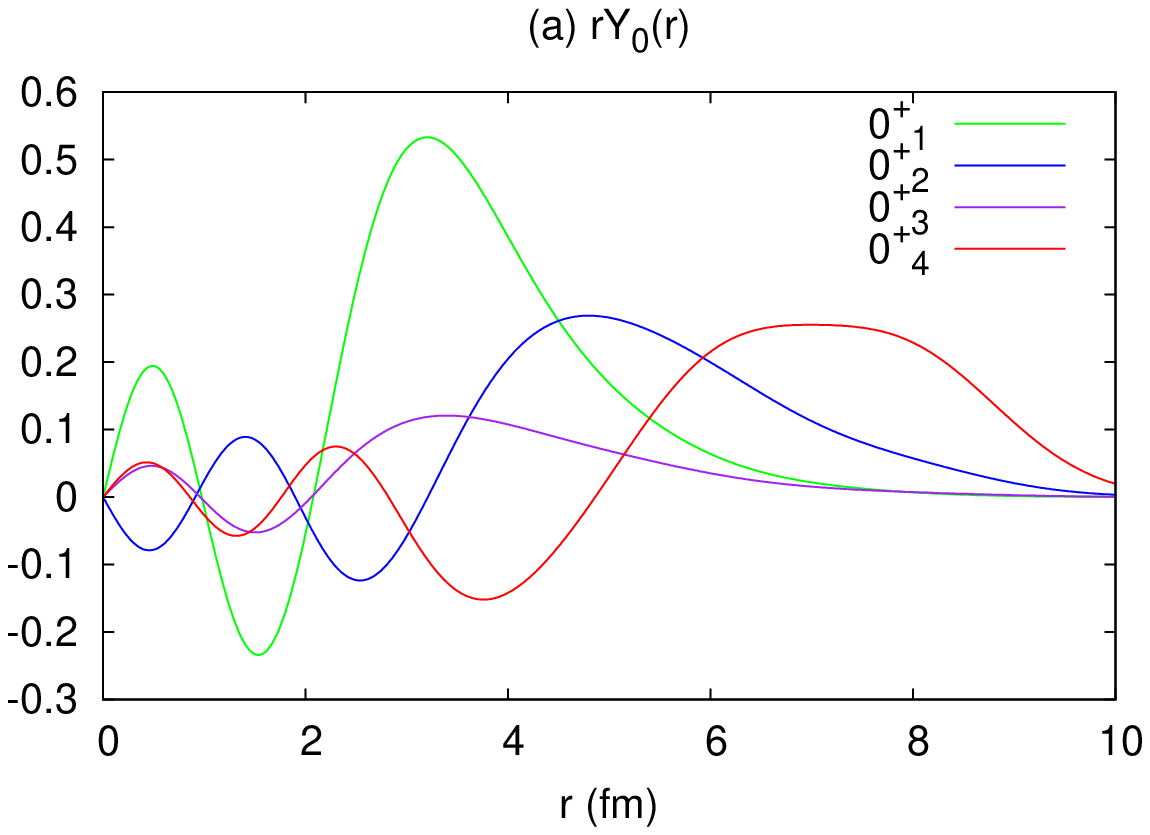}} \\
{\includegraphics[scale=0.65]{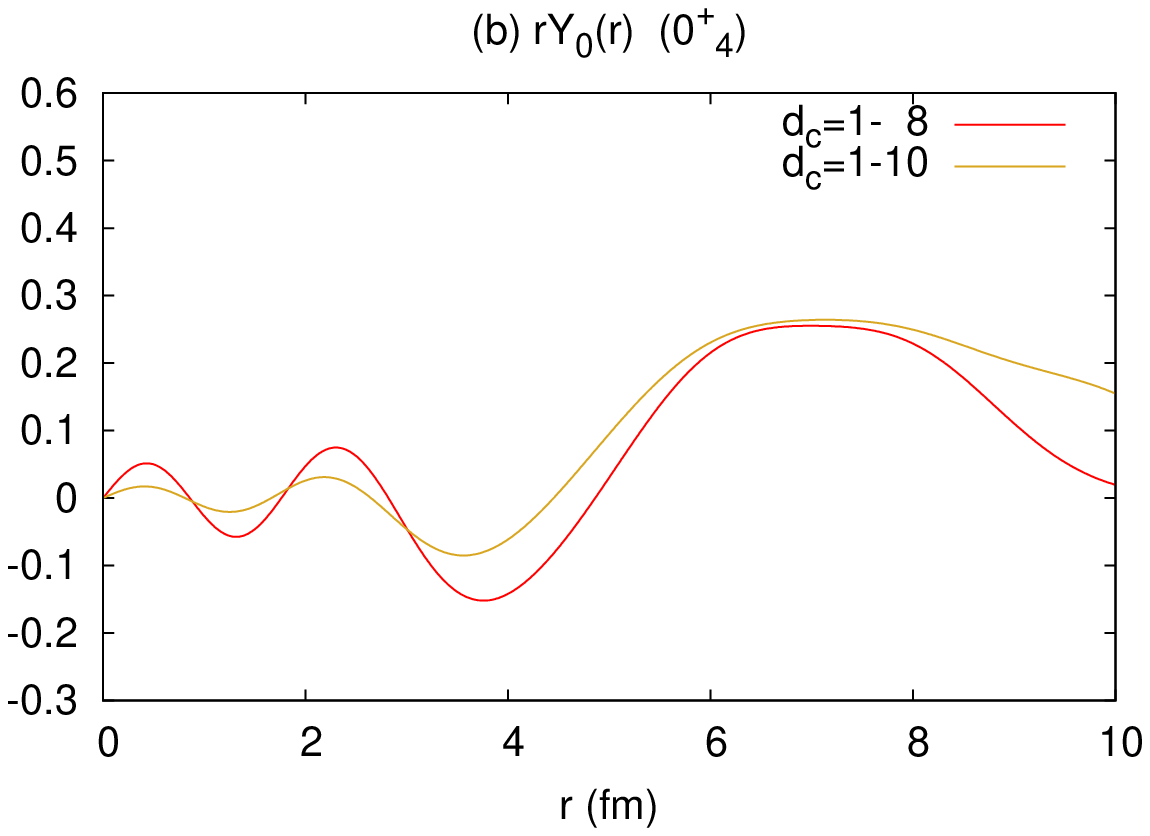}} \\
\caption{(a)The RWAs of $0^+$ states, $r\mathcal{Y}_0(r)$. 
The horizontal axis is the distance between $^6$He and $\alpha$, $r$.
(b)The RWAs of the $0^+_4$ states where 
the parameters $d_c$ of superposed $^6$He+$\alpha$ wave functions 
are up to $8$ fm or $10$ fm.}
\label{fig:RWA}
\end{figure}

\begin{table*}[t]
\caption{The $\alpha$-decay widths, $\Gamma_{\alpha}$, and 
the $\alpha$ spectroscopic factors, $S_{\alpha}$, of the $0^+_{3,4}$ states
in the cases where the resonance energy, $E_r$, is 2,3,4 MeV. 
In the calculations of the $\alpha$-decay widths, the channel radius, $a$, 
is chosen to $5,8$ fm for $0^+_{3,4}$ respectively.
$\alpha$ spectroscopic factors do not depend on the resonance energy 
so that we write the values only on $E_r = 2.0$ MeV lines.}
\label{tab:width_alpha}
\begin{ruledtabular}
\begin{tabular}{ccccccc}
&& \multicolumn{2}{c} {$d_c = 1 - 8$ fm} & 
\multicolumn{2}{c} {$d_c = 1 - 10$ fm } & \\
\hline
& $E_r$ (MeV) & $\Gamma_{\alpha}$ (keV) & $S_{\alpha}$ 
& $\Gamma_{\alpha}$ (keV) & $S_{\alpha}$ & \\
\hline
$0^+_3$ & 2.0 & $2.7 \times 10^1$ & $0.03$ & $2.8 \times 10^1$ & $0.03$ & \\
& 3.0 & $3.8 \times 10^1$ & & $4.0 \times 10^1$ & & \\
& 4.0 & $4.7 \times 10^1$ & & $4.9 \times 10^1$ & & \\
\hline
$0^+_4$ & 2.0 & $3.5 \times 10^2$ & 0.21 & $4.2 \times 10^2$ & 0.28 & \\
& 3.0 & $4.7 \times 10^2$ & & $5.6 \times 10^2$ & & \\
& 4.0 & $5.6 \times 10^2$ & & $6.6 \times 10^2$ & & \\
\end{tabular}
\end{ruledtabular}
\end{table*}

Since an $\alpha$ cluster is developed in the exotic cluster states of interest, 
especially in the $0^+_4$ state, 
they should be confirmed by surveying a decay to $^6$He and $\alpha$.
Therefore we evaluate the partial $\alpha$-decay widths
by using the reduced width amplitudes (RWAs). 
The RWAs are calculated
following an approximate method in Ref.~\cite{enyo03} 
where the eigenvalues of the norm kernel are calculated numerically. 

The RWA between $^6$He and $\alpha$, $\mathcal{Y}_L$, 
and the $\alpha$ spectroscopic factor, $S_{\alpha}$, 
are defined as follows.
\begin{gather}
\mathcal{Y}_L(a) \equiv 
\sqrt{ \frac{10!}{6!4!} } \left< \frac{\delta(r-a)}{r^2}
Y_{L0}(\hat{r}) \phi_0(^6 {\rm He}) \phi_0(\alpha) | \Phi_L \right>, \\
S_{\alpha L} \equiv \int dr \ r^2 \mathcal{Y}_L (r)^2, 
\end{gather}
where $a$ is the channel radius 
and $Y_{L0}(\hat{r})$ is the spherical harmonic function. 
$\phi_0(^6{\rm He}, \alpha)$ is the internal wave function of 
$^6$He($0^+_1$) and $\alpha$ cluster
and $\Phi_L$ is a $^{10}$Be wave function. 
Here the $^6$He($0^+_1$) wave function is the internal wave function of 
$\Phi_{^6{\rm He}(0^+_1)}$ explained in the previous section.
In the $^6$He($0^+_1$) state, 
the $(0p_{3/2})^2$ configuration of the two valence neutrons is 
dominant ($\sim 99 \%$) in the present calculation.

The calculated RWA of each $0^+$ state is shown in FIG.~\ref{fig:RWA}. 
In FIG.~\ref{fig:RWA}(a), it is seen that the $0^+_4$ state has very large amplitude 
in the large $r$ region as expected. 
Compared to the RWAs of the $0^+_1$ and $0^+_2$ states, 
the node number is the minimum in the $0^+_1$ and 
it increase one by one in the $0^+_2$ state and in the $0^+_4$ state. 
It suggests that the $0^+_4$ state might be interpreted 
as the higher nodal state of the $0^+_2$ state caused 
by the excitation of the inter-cluster motion. 
Incidentally, its amplitude is rapidly decreased over $8$ fm,  
for the present model space is truncated by $d_c = 1 - 8$ fm 
$^6$He+$\alpha$ cluster wave functions. 
As mentioned in Sec.~\ref{Sec:c+DC_wf}, 
the present $^{10}$Be wave functions are obtained within 
the bound state approximation against $^6$He+$\alpha$ continuum states.
The $0^+_4$ state actually should couple to the continuum states further
and it might have larger amplitude in the large $r$ region. 
To see the effect of further mixture of $^6$He+$\alpha$ continuum states, 
we compare the RWAs of the $0^+_4$ state calculated by superposing cluster wave functions 
with $d_c = 1 - 8$ fm 
and $d_c = 1 - 10$ fm in FIG.~\ref{fig:RWA}(b). 
In the case of $d_c = 1 - 10$ fm,  
the amplitude in the large $r$ region is actually enhanced
because of coupling with the continuum states.
However, the amplitudes around the peak at $r=7 \sim 8$ fm does not change so much, 
indicating that the present bound state is reasonable. 

Using the RWA derived as above, 
the partial $\alpha$-decay width, $\Gamma_{\alpha L}$, is estimated 
by a low-energy limit approximation \cite{lane58} as follows. 
\begin{align}
\Gamma_{\alpha L} &\ = 2P_L(ka) \gamma_{\alpha L}^2 (a), \label{eq:Gamma_alpha} \\ 
P_L(ka) &\ = \frac{ka}{F_L^2(ka) + G_L^2(ka)}, \\
\gamma_{\alpha L}^2 (a) &\ = \frac{\hbar^2}{2 \mu a}
| a \mathcal{Y}_L (a) |^2, 
\end{align}
where $\mu$ is the reduced mass of $^6$He and $\alpha$, 
and $k$ is the wave number of the resonance energy $E_r$
($k = \sqrt{2\mu E_r / \hbar^2}$). 
The partial $\alpha$ decay width, $\Gamma_{\alpha L}$, is 
the product of the reduced width, $\gamma_{\alpha L}$, 
and the penetrability, $P_L$. 
In $P_L$, 
$F_L$ and  $G_L$ are the regular and irregular Coulomb wave functions respectively. 

The $\alpha$-decay widths and 
the $\alpha$ spectroscopic factors of the $0^+_{3,4}$ states are shown
in TABLE~\ref{tab:width_alpha}. 
In the calculations of the $\alpha$-decay widths, 
we use the channel radius, $a = 5, 8$ fm for the $0^+_3$ and $0^+_4$ states respectively,
which we choose to slightly larger values than those of the main peak positions 
of the RWAs shown in FIG.~\ref{fig:RWA}(a). 
For the resonance energy, 
we use $E_r = 2,3,4$ MeV for both states, 
which are near the energy suggested in Ref.~\cite{kuchera11} 
as the resonance energy above the threshold to $^6$He+$\alpha$
for the possible $^6$He+$\alpha$ resonance state. 
The $0^+_4$ state has the larger decay width 
and $\alpha$ spectroscopic factor than those of $0^+_3$ as expected. 
When the $^6$He+$\alpha$ cluster wave functions up to $d_c = 10$ fm are superposed, 
the width and $\alpha$ spectroscopic factor are enhanced largely in the $0^+_4$ state 
because this state is very sensitive to the component of separated $^6$He and $\alpha$,
while calculated values hardly change in the $0^+_3$ state, 
which contains a little component of $^6$He+$\alpha$. 
It indicates that the $\alpha$-decay width of the $0^+_4$ state  
shown in TABLE.~\ref{tab:width_alpha} would still become larger 
by considering the coupling to continuum states fully. 
It is noted that the approximation, Eq.~(\ref{eq:Gamma_alpha}), 
is appropriate for narrow resonances 
but may be inappropriate for such broad ones as the $0^+_4$ states. 
To estimate the broad $\alpha$-decay width of the $0^+_4$the state more correctly, 
we use the formula which is the origin of the approximated Eq.~(\ref{eq:Gamma_alpha}) 
referred in Ref.~\cite{arima74}. 
\begin{equation}
\tilde{\Gamma}_{\alpha L} = 2 \left[ \frac{G_L^2}{ka} \left( \frac{1}{\gamma_{\alpha L}^2}
+ \dot{S}_L - \dot{P}_L \frac{F_L}{G_L} \right) - \dot{\phi}_L \right]^{-1}, \label{eq:Gamma_tilde}
\end{equation}
where $S$ and $\phi$ are the usual shift function 
and the hard-sphere scattering phase shift, 
and the dots signify the energy derivative. 
Applying this formula to the $0^+_4$ state
formed by superposing $d_c = 1 - 10$ fm $^6$He+$\alpha$ cluster wave functions, 
we obtain $\tilde{\Gamma}_{\alpha} = 8.2 \times 10^2$ keV
for $E_r = 3.0$ MeV and $a = 8.0$ fm, 
which is about 50 \% larger than the $\Gamma_{\alpha}$ evaluated with Eq.(\ref{eq:Gamma_alpha}).

Besides, although we show here the results of the representative states 
having the major component of each state, 
part of the components of the $0^+_{3,4}$ states are 
fragmented in neighboring continuum states. 
Therefore, the actual $\alpha$-decay width
which is the sum of those fragments 
would be somewhat larger than our calculated values. 
It should be also noticed that we do not consider a one- and two-neutron decays,
and therefore, the total decay width should be still larger.
In order to calculate the more accurate values for the widths, 
it is essential to perform a more sophisticated treatment of a resonance with a broad width
considering all the decay channels, 
but it is beyond the present study. 
We just emphasize that the $0^+_4$ state certainly has the large $\alpha$-decay width 
and it can be confirmed by surveying $\alpha$-decay.

\subsubsection{The monopole transition strength}
\label{Sec:monopole}
\begin{table*}[t]
\caption{The total, proton and neutron monopole transition strength 
from the $0^+_{\nu}$ state to the ground state, $M^{t,p,n}_{\nu}$.
In the parentheses, the ratios to the first-order EWSR, $S^{t,p,n}_1$, 
are shown by \%.}
\label{tab:monopole}
\begin{ruledtabular}
\begin{tabular}{ccccccc}
& $M^t_{\nu}$ && $M^p_{\nu}$ && $M^n_{\nu}$ & \\
\hline
$0^+_2$ & 
5.9 & \hspace{-5em} ( 5.3 \% ) & 
2.1 & \hspace{-5em} ( 1.8 \% ) & 
3.8 & \hspace{-5em} ( 3.5 \% ) \\
$0^+_3$ & 
8.6 & \hspace{-5em} ( 13.6 \% ) & 
1.5 & \hspace{-5em} ( 1.1 \% ) & 
7.1 & \hspace{-5em} ( 14.7 \% ) \\
$0^+_4$ & 
2.9 & \hspace{-5em} ( 2.2 \% ) & 
2.4 & \hspace{-5em} ( 4.1 \% ) & 
0.5 & \hspace{-5em} ( 8.8 $\times 10^{-2}$ \% ) \\
\end{tabular}
\end{ruledtabular}
\end{table*}

The monopole transition is expected to 
be a measure of the extension of clusters \cite{yamada08,yamada08_2}. 
Two factors are essential for a large monopole transition strength. 
The first is that states before and/or after transition have a extended structure 
because the monopole transition operator is composed of $r^2$ (Eq.~(\ref{eq:monopole}))
so that it reflects the degree of a radial excitation. 
The second is that the configuration does not so differ before and after transition, 
for example, it can be large in the case of the transition from an excited cluster state to
a ground state which involves some cluster components. 
So it is expected that a remarkable monopole transition strength can be observed in $^{10}$Be 
where the ground and excited states have striking cluster structures. 

The total monopole transition strength 
between the excited $0^+_{\nu}$ states and the ground state, $M^t_{\nu}$, 
is defined as 
\begin{equation}
M^t_{\nu} = \langle 0^+_{\nu}|\sum_i r_i^2|0^+_1 \rangle, \label{eq:monopole}
\end{equation}
where the index $i$ runs over all of the nucleons. 
The proton and neutron ones, $M^{p,n}_{\nu}$, are defined in the same way, 
though the index $i$ runs over only protons or neutrons.
Here these calculations are performed
by superposing DC wave functions whose $\beta$ is reduced to $2 - 6$ fm
and $^6$He+$\alpha$ cluster wave functions with $d_c = 1 - 8$ fm. 
Both of the $0^+_{3,4}$ states are spatially extended quasi-bound states
so that they tend to be mixed with continuum states 
of $\alpha$+$\alpha$+$n$+$n$ and $^6$He+$\alpha$ respectively. 
The more continuum states are coupled, 
the more strength increases due to their contributions. 
Therefore, with the inclusion of a larger model space and more coupling with continuum state, 
calculated monopole transition strengths become very uncertainty. 
So in the calculations of monopole transition strengths, 
we adopt the further restricted DC wave functions.
We have confirmed that 
their qualitative structures do not differ from those of the original ones.
As the criterion of a degree of the strength, 
we also calculate the ratios to the first-order energy-weighted sum rule (EWSR), 
\begin{align}
S^t_1 = &\ \sum_{\nu} ( E_{\nu} - E_0 ) 
|\langle 0^+_{\nu}|\sum_i r_i^2|0^+_1 \rangle|^2, \\
= &\ \frac{2 \hbar^2}{m} \langle 0^+_1|\sum_i r_i^2|0^+_1 \rangle, 
\label{eq:EWSR}
\end{align}
where $E_{\nu}-E_0$ is the excited energy of the $\nu$th $0^+$ state. 
The second line, Eq.~(\ref{eq:EWSR}), holds 
in the case that the Hamiltonian contains only momentum-independent interactions.
The definitions of the superscripts $t,p,n$ are the same 
as those in the monopole transition strength. 
The calculated values of the strengths between the ground state and the $0^+_{2,3,4}$ states
and their ratios to the EWSR are shown in TABLE~\ref{tab:monopole}. 
The $0^+_3$ state has a remarkable strength in the neutron one 
and it exhausts 14.7 \% of the EWSR. 
It reflects the remarkable gas-like structure of the $0^+_3$ state shown 
in FIG.~\ref{fig:overlap_B_b}. 
This state has two $\alpha$ and one dineutron clusters 
which are weakly interacting with each other
so that the neutrons are spatially developed largely. 
Moreover, the ground state contains two $\alpha$ clusters and remarkable dineutron correlation 
at the surface as shown in Ref.~\cite{kobayashi11}.
As a result, its monopole transition strength is highly enhanced.
In contrast to the $0^+_3$, 
the $0^+_4$ state has just a little strength of a few \% of the EWSR. 
It is because the $0^+_4$ has the developed $\alpha$ cluster structure
which is so extended that the overlap with the ground state is too small. 
Indeed, the RWA of the $0^+_4$ has the second higher nodal structure of 
that of the $0^+_1$ state 
and shows different behaviors (FIG.~\ref{fig:RWA}(a)).
Thus the distinct structures between the $0^+_3$ and $0^+_4$ states cause 
the clear difference in their monopole transition strengths, 
which would be a good probe to identify the $0^+_3$ and $0^+_4$ states.
It is indeed meaningful that 
not the absolute values but the relative magnitude
since we superpose only restricted wave functions
in the present calculation of the monopole transition strengths. 
The present result suggests that 
the $0^+_3$ state has the more striking monopole transition strength 
than those of the other $0^+$ states,
so that we expect that the $0^+_3$ state can be confirmed experimentally 
in the inelastic scattering from the ground state.

\section{Summary}
\label{Sec:summary}

We have investigated the exotic cluster structures of excited states of $^{10}$Be by using the $^6$He+$\alpha$ cluster wave functions plus the DC wave functions. 
In the present study, we suggest theoretically two kinds of novel cluster states 
above the $\alpha$+$\alpha$+$n$+$n$ threshold energy, 
which have not been confirmed experimentally yet. 
These states have remarkable cluster structures 
and construct the $K^{\pi} = 0^+$ rotational bands 
which show quite distinct characteristics. 
Ones of them contain developed two $\alpha$ and one dineutron clusters. 
They are weakly interacting to each other, 
and the $0^+$ state in the band is regarded as a gas-like structure in association with 
the Hoyle state, $^{12}$C($0^+_2$). 
In addition to such states, 
we also found states composed of $^6$He and a developed $\alpha$ cluster. 
Because of a large $\alpha$ spectroscopic factor, 
the $0^+$ state is likely to correspond to the one suggested experimentally 
on the analogy of $^{10}$B. 
In this state, the $\alpha$ cluster is extremely extended from $^6$He 
so that one could even call it $\alpha$-halo. 

Since these new states have distinct cluster structures,  
they should be distinguished with the different observables.
We have suggested that 
the $0^+$ state including extended two $\alpha$s and one dineutron 
can have a large monopole transition strength from the ground state 
due to the gas-like extended $\alpha$+$\alpha$+dineutron structure. 
On the other hand, 
the $^6$He+$\alpha$ state has an extremely developed $\alpha$ cluster
so that it can be confirmed by measuring the decay to $^6$He and $\alpha$. 
We have made sure that the $^6$He+$\alpha$ state has
a much larger $\alpha$-decay width than the $\alpha$+$\alpha$+dineutron state. 

In the present work, we have mainly focused on 
the structures of two novel cluster $0^+$ states of $^{10}$Be. 
These states are obtained within the bound state approximation 
and are suggested to be resonance states 
above the $\alpha$+$\alpha$+$n$+$n$ and $^6$He+$\alpha$ threshold energies. 
However, in order to predict precise resonance energies and widths of these states,  
further improvements of the calculation are required. 
For instance, 
we need to treat resonance behaviors beyond the bound state approximations 
by taking into account all the decay channel which are neglected here 
and coupling with continuum states. 
Moreover, the threshold energies, in particular, 
those of $\alpha$+$\alpha$+$n$+$n$ and $^6$He+$\alpha$ channels should be reproduced 
by improving wave functions and adjusting effective interactions carefully. 
The purpose of this work is to suggest exotic cluster states 
and the points mentioned here are tasks for the future.  

Dineutron correlation is usually discussed in study of ground state properties. 
What we have shown in this paper is that
it can be important not only in ground states but also in excited states. 
As a future work, we will apply our method systematically 
to the nuclei neighboring $^{10}$Be, that is, $^9$Li and $^8$He, 
and search for states containing one or a few dineutron clusters 
and investigate the tendency of dineutron cluster formation in those nuclei.

\appendix*
\section{Identification of the resonance states}
\label{appendix}
Here we show the additional way to identify the resonance states in continuum spectra
which we assign to the members of the $K^{\pi} = 0^+$ bands
labeled as $0^+_{3,4}$ and so on in this paper. 

Since we superpose a number of wave functions 
to describe the quasi-bound cluster states, 
there appear a number of continuum states in addition to the resonance states.  
As we describe in Sec.~\ref{Sec:result}, 
we identify the $0^+_{3,4}$ states 
by means of the significant amplitude of $\alpha$+$\alpha$+dineutron 
and $^6$He+$\alpha$ in a finite region respectively. 
However, these criteria may seem to be ambiguous. 
To make sure the identification further, 
we perform additional analysis following the pseudo potential method done in Ref.~\cite{enyo12}. 

In this method, we introduce a pseudo potential, $\tilde{V}$, 
and add it to the original Hamiltonian, $H$, (Eq. (\ref{eq:hamiltonian})), 
\begin{gather}
\tilde{H}(\delta) = H + \delta \times \tilde{V}, 
\label{eq:pseudo_pot} \\
\tilde{V} = \sum_{i<j} v_0 \exp \left[ - \frac{r_{ij}^2}{a_0^2} \right], 
\end{gather}
where as the parameters $v_0 = -100$ MeV and $a_0 = 1.0$ fm are chosen. 
$\delta$ is the parameter to control the strength of the pseudo potential.
When $\delta = 0$, the modified Hamiltonian, $\tilde{H}(\delta)$, 
defined in Eq.~(\ref{eq:pseudo_pot}) is equal to the original one, $H$. 
When $\delta$ is increased, 
since the short-range attraction between nucleons becomes artificially larger, 
resonance states which have larger components in a finite region 
should gain relatively more energy than continuum states. 
With an enough large $\delta$, 
the resonance states come down below continuum states to become bound states. 
As a result, the states of interest can be separated out of continuum states energetically. 

\begin{figure}[b]
\includegraphics[scale=0.65]{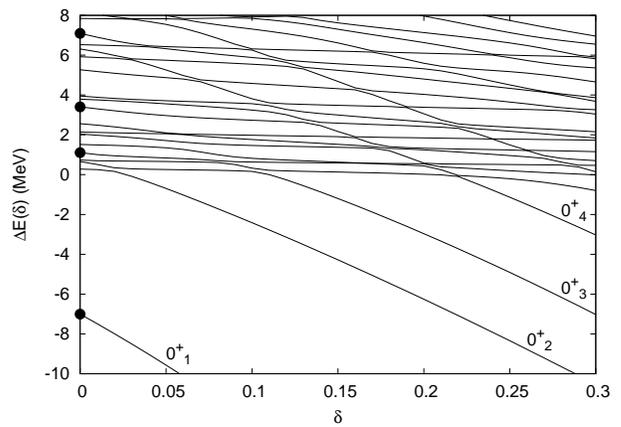}
\caption{The transition of the energies of the $0^+$ states as the pseudo potential is increased. 
The horizontal axis is the parameter of the strength of the pseudo potential, $\delta$.
The points at $\delta = 0$ correspond to the $0^+$ states plotted in FIG.~\ref{fig:ene_spect}.}
\label{fig:ene_ACCC}
\end{figure}

By diagonalizing the Hamiltonian matrix calculated with $\tilde{H}(\delta)$, 
energy levels are obtained. 
The energies of $0^+$ states calculated as below are plotted in FIG.~\ref{fig:ene_ACCC}. 
\begin{equation}
\Delta E (\delta) \equiv \langle \Psi^{0^+}(\delta)| \tilde{H}(\delta) |\Psi^{0^+}(\delta) \rangle
- \langle \Phi_{2\alpha}^{0^+}(\delta)| \tilde{H}(\delta) |\Phi_{2\alpha}^{0^+}(\delta) \rangle. 
\end{equation}
Here we show the energies measured from the one of the 2$\alpha$ core
calculated by superposing the same basis wave functions as the core in Eq.~(\ref{eq:DCwf}), 
that is, two $\alpha$s whose distances are $1 - 6$ fm, 
and projected it to $J^{\pi} = 0^+$. 
In FIG.~\ref{fig:ene_ACCC}, 
it is shown that energies for most of states are almost independent of the parameter $\delta$ 
as shown by plateau lines in FIG.~\ref{fig:ene_ACCC}. 
These states may correspond to continuum states. 
Among those continuum states, there exist resonance states whose energy decreases significantly 
as the parameter $\delta$ increases. 
When $\delta$ is large enough, 
the resonance states are decoupled from continuum states thoroughly 
to form bound states. 
In such a decoupling region, the $0^+_3$ and $0^+_4$ states are composed 
of the prominent components of $\alpha$+$\alpha$+dineutron and $^6$He+$\alpha$, respectively. 
The corresponding states at $\delta = 0$ (the points at $\delta = 0$ in FIG.~\ref{fig:ene_ACCC}) 
are consistent with the assignment of the $0^+_3$ and $0^+_4$ states 
discussed in Sec.~\ref{Sec:result}. 

In the present calculation, main continuum states are 2$\alpha$ core plus free two neutrons 
due to the present bound state approximation 
where $^6$He+$\alpha$ continuum states are not considered sufficiently. 
$^6$He+$\alpha$ continuum states can be included 
by superposing $^6$He+$\alpha$ cluster wave functions with larger $d_c$. 
When ones up to $d_c = 15$ fm are superposed, 
some $^6$He+$\alpha$ continuum states appear below the $0^+_4$ state. 
We have applied the pseudo potential method with such bases 
and confirmed that the resonance states are decoupled from 
both continuum states of 2$\alpha$+$n$+$n$ and $^6$He+$\alpha$ 
in the same manner in the above analysis. 

To assign the other members of their rotational bands, 
we have also applied the same method with respect to $2^+$ and $4^+$ states.

\begin{acknowledgments}
This work was supported by Grant-in-Aid for Scientific Research 
from Japan Society for the Promotion of Science (JSPS).
It was also supported by
the Grant-in-Aid for the Global COE Program ``The Next Generation of Physics,
Spun from Universality and Emergence'' 
from the Ministry of Education, Culture, Sports, Science and Technology (MEXT) of Japan.
A part of the computational calculations of this work was performed by using the
supercomputers at YITP.
\end{acknowledgments}

\bibliography{reference}

\begin{thebibliography}{28}
\expandafter\ifx\csname natexlab\endcsname\relax\def\natexlab#1{#1}\fi
\expandafter\ifx\csname bibnamefont\endcsname\relax
  \def\bibnamefont#1{#1}\fi
\expandafter\ifx\csname bibfnamefont\endcsname\relax
  \def\bibfnamefont#1{#1}\fi
\expandafter\ifx\csname citenamefont\endcsname\relax
  \def\citenamefont#1{#1}\fi
\expandafter\ifx\csname url\endcsname\relax
  \def\url#1{\texttt{#1}}\fi
\expandafter\ifx\csname urlprefix\endcsname\relax\def\urlprefix{URL }\fi
\providecommand{\bibinfo}[2]{#2}
\providecommand{\eprint}[2][]{\url{#2}}

\bibitem[{\citenamefont{Seya et~al.}(1981)\citenamefont{Seya, Kohno, and
  Nagata}}]{seya81}
\bibinfo{author}{\bibfnamefont{M.}~\bibnamefont{Seya}},
  \bibinfo{author}{\bibfnamefont{M.}~\bibnamefont{Kohno}}, \bibnamefont{and}
  \bibinfo{author}{\bibfnamefont{S.}~\bibnamefont{Nagata}},
  \bibinfo{journal}{Prog.~Theor.~Phys.} \textbf{\bibinfo{volume}{65}},
  \bibinfo{pages}{204} (\bibinfo{year}{1981}).

\bibitem[{\citenamefont{von Oertzen}(1996)}]{oertzen96}
\bibinfo{author}{\bibfnamefont{W.}~\bibnamefont{von Oertzen}},
  \bibinfo{journal}{Z.~Phys.~A} \textbf{\bibinfo{volume}{354}},
  \bibinfo{pages}{37} (\bibinfo{year}{1996}).

\bibitem[{\citenamefont{Arai et~al.}(1996)\citenamefont{Arai, Ogawa, Suzuki,
  and Varga}}]{arai96}
\bibinfo{author}{\bibfnamefont{K.}~\bibnamefont{Arai}},
  \bibinfo{author}{\bibfnamefont{Y.}~\bibnamefont{Ogawa}},
  \bibinfo{author}{\bibfnamefont{Y.}~\bibnamefont{Suzuki}}, \bibnamefont{and}
  \bibinfo{author}{\bibfnamefont{K.}~\bibnamefont{Varga}},
  \bibinfo{journal}{Phys.~Rev.~C} \textbf{\bibinfo{volume}{54}},
  \bibinfo{pages}{132} (\bibinfo{year}{1996}).

\bibitem[{\citenamefont{Kanada-En'yo et~al.}(1999)\citenamefont{Kanada-En'yo,
  Horiuchi, and Dot\'{e}}}]{enyo99}
\bibinfo{author}{\bibfnamefont{Y.}~\bibnamefont{Kanada-En'yo}},
  \bibinfo{author}{\bibfnamefont{H.}~\bibnamefont{Horiuchi}}, \bibnamefont{and}
  \bibinfo{author}{\bibfnamefont{A.}~\bibnamefont{Dot\'{e}}},
  \bibinfo{journal}{Phys.~Rev.~C} \textbf{\bibinfo{volume}{60}},
  \bibinfo{pages}{064304} (\bibinfo{year}{1999}).

\bibitem[{\citenamefont{Itagaki and Okabe}(2000)}]{itagaki00}
\bibinfo{author}{\bibfnamefont{N.}~\bibnamefont{Itagaki}} \bibnamefont{and}
  \bibinfo{author}{\bibfnamefont{S.}~\bibnamefont{Okabe}},
  \bibinfo{journal}{Phys.~Rev.~C} \textbf{\bibinfo{volume}{61}},
  \bibinfo{pages}{044306} (\bibinfo{year}{2000}).

\bibitem[{\citenamefont{Ogawa et~al.}(2000)\citenamefont{Ogawa, Arai, Suzuki,
  and Varga}}]{ogawa00}
\bibinfo{author}{\bibfnamefont{Y.}~\bibnamefont{Ogawa}},
  \bibinfo{author}{\bibfnamefont{K.}~\bibnamefont{Arai}},
  \bibinfo{author}{\bibfnamefont{Y.}~\bibnamefont{Suzuki}}, \bibnamefont{and}
  \bibinfo{author}{\bibfnamefont{K.}~\bibnamefont{Varga}},
  \bibinfo{journal}{Nucl.~Phys.~A} \textbf{\bibinfo{volume}{673}},
  \bibinfo{pages}{122} (\bibinfo{year}{2000}).

\bibitem[{\citenamefont{Kanada-En'yo and Horiuchi}(2003)}]{enyo03}
\bibinfo{author}{\bibfnamefont{Y.}~\bibnamefont{Kanada-En'yo}}
  \bibnamefont{and} \bibinfo{author}{\bibfnamefont{H.}~\bibnamefont{Horiuchi}},
  \bibinfo{journal}{Phys.~Rev.~C} \textbf{\bibinfo{volume}{68}},
  \bibinfo{pages}{014319} (\bibinfo{year}{2003}).

\bibitem[{\citenamefont{Ito et~al.}(2004)\citenamefont{Ito, Kato, and
  Ikeda}}]{ito04}
\bibinfo{author}{\bibfnamefont{M.}~\bibnamefont{Ito}},
  \bibinfo{author}{\bibfnamefont{K.}~\bibnamefont{Kato}}, \bibnamefont{and}
  \bibinfo{author}{\bibfnamefont{K.}~\bibnamefont{Ikeda}},
  \bibinfo{journal}{Phys.~Lett.~B} \textbf{\bibinfo{volume}{588}},
  \bibinfo{pages}{43} (\bibinfo{year}{2004}).

\bibitem[{\citenamefont{Ito}(2006)}]{ito06}
\bibinfo{author}{\bibfnamefont{M.}~\bibnamefont{Ito}},
  \bibinfo{journal}{Phys.~Lett.~B} \textbf{\bibinfo{volume}{636}},
  \bibinfo{pages}{293} (\bibinfo{year}{2006}).

\bibitem[{\citenamefont{Suhara and Kanada-En'yo}(2010)}]{suhara10}
\bibinfo{author}{\bibfnamefont{T.}~\bibnamefont{Suhara}} \bibnamefont{and}
  \bibinfo{author}{\bibfnamefont{Y.}~\bibnamefont{Kanada-En'yo}},
  \bibinfo{journal}{Prog.~Theor.~Phys.} \textbf{\bibinfo{volume}{123}},
  \bibinfo{pages}{303} (\bibinfo{year}{2010}).

\bibitem[{\citenamefont{Kuchera et~al.}(2011)\citenamefont{Kuchera, Rogachev,
  Goldberg, Johnson, Cherubini, Gulino, Cognata, Lamia, Romano, Miller
  et~al.}}]{kuchera11}
\bibinfo{author}{\bibfnamefont{N.}~\bibnamefont{Kuchera}},
  \bibinfo{author}{\bibfnamefont{G.~V.} \bibnamefont{Rogachev}},
  \bibinfo{author}{\bibfnamefont{V.~Z.} \bibnamefont{Goldberg}},
  \bibinfo{author}{\bibfnamefont{E.~D.} \bibnamefont{Johnson}},
  \bibinfo{author}{\bibfnamefont{S.}~\bibnamefont{Cherubini}},
  \bibinfo{author}{\bibfnamefont{M.}~\bibnamefont{Gulino}},
  \bibinfo{author}{\bibfnamefont{M.~L.} \bibnamefont{Cognata}},
  \bibinfo{author}{\bibfnamefont{L.}~\bibnamefont{Lamia}},
  \bibinfo{author}{\bibfnamefont{S.}~\bibnamefont{Romano}},
  \bibinfo{author}{\bibfnamefont{L.~E.} \bibnamefont{Miller}},
  \bibnamefont{et~al.}, \bibinfo{journal}{Phys.~Rev.~C}
  \textbf{\bibinfo{volume}{84}}, \bibinfo{pages}{054615}
  (\bibinfo{year}{2011}).

\bibitem[{\citenamefont{Baldo et~al.}(1990)\citenamefont{Baldo, Cugnon,
  Lejeune, and Lombardo}}]{baldo90}
\bibinfo{author}{\bibfnamefont{M.}~\bibnamefont{Baldo}},
  \bibinfo{author}{\bibfnamefont{J.}~\bibnamefont{Cugnon}},
  \bibinfo{author}{\bibfnamefont{A.}~\bibnamefont{Lejeune}}, \bibnamefont{and}
  \bibinfo{author}{\bibfnamefont{U.}~\bibnamefont{Lombardo}},
  \bibinfo{journal}{Nucl.~Phys.~A} \textbf{\bibinfo{volume}{515}},
  \bibinfo{pages}{409} (\bibinfo{year}{1990}).

\bibitem[{\citenamefont{Matsuo}(2006)}]{matsuo06}
\bibinfo{author}{\bibfnamefont{M.}~\bibnamefont{Matsuo}},
  \bibinfo{journal}{Phys.~Rev.~C} \textbf{\bibinfo{volume}{73}},
  \bibinfo{pages}{044309} (\bibinfo{year}{2006}).

\bibitem[{\citenamefont{Bertsch and Esbensen}(1991)}]{bertsch91}
\bibinfo{author}{\bibfnamefont{G.~F.} \bibnamefont{Bertsch}} \bibnamefont{and}
  \bibinfo{author}{\bibfnamefont{H.}~\bibnamefont{Esbensen}},
  \bibinfo{journal}{Ann.~Phys.~} \textbf{\bibinfo{volume}{209}},
  \bibinfo{pages}{327} (\bibinfo{year}{1991}).

\bibitem[{\citenamefont{Zhukov et~al.}(1993)\citenamefont{Zhukov, Danilin,
  Fedrov, Bang, Thompson, and Vaagen}}]{zhukov93}
\bibinfo{author}{\bibfnamefont{M.~V.} \bibnamefont{Zhukov}},
  \bibinfo{author}{\bibfnamefont{B.~V.} \bibnamefont{Danilin}},
  \bibinfo{author}{\bibfnamefont{D.~V.} \bibnamefont{Fedrov}},
  \bibinfo{author}{\bibfnamefont{J.~M.} \bibnamefont{Bang}},
  \bibinfo{author}{\bibfnamefont{I.~J.} \bibnamefont{Thompson}},
  \bibnamefont{and} \bibinfo{author}{\bibfnamefont{J.~S.}
  \bibnamefont{Vaagen}}, \bibinfo{journal}{Phys.~Rep.}
  \textbf{\bibinfo{volume}{231}}, \bibinfo{pages}{151} (\bibinfo{year}{1993}).

\bibitem[{\citenamefont{Matsuo et~al.}(2005)\citenamefont{Matsuo, Mizuyama, and
  Serizawa}}]{matsuo05}
\bibinfo{author}{\bibfnamefont{M.}~\bibnamefont{Matsuo}},
  \bibinfo{author}{\bibfnamefont{K.}~\bibnamefont{Mizuyama}}, \bibnamefont{and}
  \bibinfo{author}{\bibfnamefont{Y.}~\bibnamefont{Serizawa}},
  \bibinfo{journal}{Phys. Rev. C} \textbf{\bibinfo{volume}{71}},
  \bibinfo{pages}{064326} (\bibinfo{year}{2005}).

\bibitem[{\citenamefont{Hagino and Sagawa}(2005)}]{hagino05}
\bibinfo{author}{\bibfnamefont{K.}~\bibnamefont{Hagino}} \bibnamefont{and}
  \bibinfo{author}{\bibfnamefont{H.}~\bibnamefont{Sagawa}},
  \bibinfo{journal}{Phys.~Rev.~C} \textbf{\bibinfo{volume}{72}},
  \bibinfo{pages}{044321} (\bibinfo{year}{2005}).

\bibitem[{\citenamefont{Kanada-En'yo}(2007)}]{enyo07}
\bibinfo{author}{\bibfnamefont{Y.}~\bibnamefont{Kanada-En'yo}},
  \bibinfo{journal}{Phys.~Rev.~C} \textbf{\bibinfo{volume}{76}},
  \bibinfo{pages}{044323} (\bibinfo{year}{2007}).

\bibitem[{\citenamefont{Kobayashi and Kanada-En'yo}(2011)}]{kobayashi11}
\bibinfo{author}{\bibfnamefont{F.}~\bibnamefont{Kobayashi}} \bibnamefont{and}
  \bibinfo{author}{\bibfnamefont{Y.}~\bibnamefont{Kanada-En'yo}},
  \bibinfo{journal}{Prog.~Theor.~Phys.} \textbf{\bibinfo{volume}{126}},
  \bibinfo{pages}{457} (\bibinfo{year}{2011}).

\bibitem[{\citenamefont{Itagaki et~al.}(2008)\citenamefont{Itagaki, Ito, Arai,
  Aoyama, and Kokalova}}]{itagaki08}
\bibinfo{author}{\bibfnamefont{N.}~\bibnamefont{Itagaki}},
  \bibinfo{author}{\bibfnamefont{M.}~\bibnamefont{Ito}},
  \bibinfo{author}{\bibfnamefont{K.}~\bibnamefont{Arai}},
  \bibinfo{author}{\bibfnamefont{S.}~\bibnamefont{Aoyama}}, \bibnamefont{and}
  \bibinfo{author}{\bibfnamefont{T.}~\bibnamefont{Kokalova}},
  \bibinfo{journal}{Phys.~Rev.~C} \textbf{\bibinfo{volume}{78}},
  \bibinfo{pages}{017306} (\bibinfo{year}{2008}).

\bibitem[{\citenamefont{Kanada-En'yo and Suhara}(2012)}]{enyo12}
\bibinfo{author}{\bibfnamefont{Y.}~\bibnamefont{Kanada-En'yo}}
  \bibnamefont{and} \bibinfo{author}{\bibfnamefont{T.}~\bibnamefont{Suhara}},
  \bibinfo{journal}{Phys.~Rev.~C} \textbf{\bibinfo{volume}{85}},
  \bibinfo{pages}{024303} (\bibinfo{year}{2012}).

\bibitem[{\citenamefont{Volkov}(1965)}]{volkov65}
\bibinfo{author}{\bibfnamefont{A.}~\bibnamefont{Volkov}},
  \bibinfo{journal}{Nuc.~Phys.} \textbf{\bibinfo{volume}{74}},
  \bibinfo{pages}{33} (\bibinfo{year}{1965}).

\bibitem[{\citenamefont{Tamagaki}(1968)}]{tamagaki68}
\bibinfo{author}{\bibfnamefont{R.}~\bibnamefont{Tamagaki}},
  \bibinfo{journal}{Prog.~Theor.~Phys.} \textbf{\bibinfo{volume}{39}},
  \bibinfo{pages}{91} (\bibinfo{year}{1968}).

\bibitem[{\citenamefont{Funaki et~al.}(2005)\citenamefont{Funaki, Tohsaki,
  Horiuchi, Schuck, and R{\"{o}}pke}}]{funaki05}
\bibinfo{author}{\bibfnamefont{Y.}~\bibnamefont{Funaki}},
  \bibinfo{author}{\bibfnamefont{A.}~\bibnamefont{Tohsaki}},
  \bibinfo{author}{\bibfnamefont{H.}~\bibnamefont{Horiuchi}},
  \bibinfo{author}{\bibfnamefont{P.}~\bibnamefont{Schuck}}, \bibnamefont{and}
  \bibinfo{author}{\bibfnamefont{G.}~\bibnamefont{R{\"{o}}pke}},
  \bibinfo{journal}{Eur.~Phys.~J.~A} \textbf{\bibinfo{volume}{24}},
  \bibinfo{pages}{321} (\bibinfo{year}{2005}).

\bibitem[{\citenamefont{Lane and Thomas}(1958)}]{lane58}
\bibinfo{author}{\bibfnamefont{A.~M.} \bibnamefont{Lane}} \bibnamefont{and}
  \bibinfo{author}{\bibfnamefont{R.~G.} \bibnamefont{Thomas}},
  \bibinfo{journal}{Rev.~Mod.~Phys.} \textbf{\bibinfo{volume}{30}},
  \bibinfo{pages}{257} (\bibinfo{year}{1958}).

\bibitem[{\citenamefont{Arima and Yoshida}(1974)}]{arima74}
\bibinfo{author}{\bibfnamefont{A.}~\bibnamefont{Arima}} \bibnamefont{and}
  \bibinfo{author}{\bibfnamefont{S.}~\bibnamefont{Yoshida}},
  \bibinfo{journal}{Nucl.~Phys.~A} \textbf{\bibinfo{volume}{219}},
  \bibinfo{pages}{475} (\bibinfo{year}{1974}).

\bibitem[{\citenamefont{Yamada et~al.}(2008{\natexlab{a}})\citenamefont{Yamada,
  Horiuchi, Ikeda, Funaki, and Tohsaki}}]{yamada08}
\bibinfo{author}{\bibfnamefont{T.}~\bibnamefont{Yamada}},
  \bibinfo{author}{\bibfnamefont{H.}~\bibnamefont{Horiuchi}},
  \bibinfo{author}{\bibfnamefont{K.}~\bibnamefont{Ikeda}},
  \bibinfo{author}{\bibfnamefont{Y.}~\bibnamefont{Funaki}}, \bibnamefont{and}
  \bibinfo{author}{\bibfnamefont{A.}~\bibnamefont{Tohsaki}},
  \bibinfo{journal}{Journal of Physics} \textbf{\bibinfo{volume}{111}},
  \bibinfo{pages}{012008} (\bibinfo{year}{2008}{\natexlab{a}}).

\bibitem[{\citenamefont{Yamada et~al.}(2008{\natexlab{b}})\citenamefont{Yamada,
  Funaki, Horiuchi, Ikeda, and Tohsaki}}]{yamada08_2}
\bibinfo{author}{\bibfnamefont{T.}~\bibnamefont{Yamada}},
  \bibinfo{author}{\bibfnamefont{Y.}~\bibnamefont{Funaki}},
  \bibinfo{author}{\bibfnamefont{H.}~\bibnamefont{Horiuchi}},
  \bibinfo{author}{\bibfnamefont{K.}~\bibnamefont{Ikeda}}, \bibnamefont{and}
  \bibinfo{author}{\bibfnamefont{A.}~\bibnamefont{Tohsaki}},
  \bibinfo{journal}{Prog.~Theor.~Phys.} \textbf{\bibinfo{volume}{120}},
  \bibinfo{pages}{1139} (\bibinfo{year}{2008}{\natexlab{b}}).

\end{thebibliography}

\end{document}